\documentclass{aa}  

\usepackage{graphicx}
\usepackage{epsfig}
\usepackage{epstopdf}
\usepackage{txfonts}
\usepackage{natbib}
\usepackage{url}
\usepackage{amssymb}
\newcommand{\raiseChi}

\begin{document}

\title{A submillimeter search for pre- and proto-brown dwarfs in Chamaeleon II}
   \author{
         I. de Gregorio-Monsalvo\inst{\ref{inst1},\ref{inst2}} 
        \and
        D. Barrado\inst{\ref{inst3}}
          \and
       H. Bouy\inst{\ref{inst3}}
          \and 
     A. Bayo\inst{\ref{inst4},\ref{inst5}}
        \and 
      Aina Palau\inst{\ref{inst6}}
          \and 
      M. Morales-Calder\'on\inst{\ref{inst3}}
         \and 
       N. Hu\'elamo\inst{\ref{inst3}}  
          \and
      O. Morata\inst{\ref{inst7}}   
            \and 
      B. Mer\'in\inst{\ref{inst8}} 
        \and 
      C. Eiroa\inst{\ref{inst9}}   
           }     
    \offprints{I. de Gregorio-Monsalvo}
  
\institute
{European Southern Observatory, Karl Schwarzschild Str 2, D-85748 Garching bei M\"unchen, Germany \email{idegrego@eso.org}\label{inst1}   
\and 
Joint ALMA office, Alonso de C\'ordova 3107, Vitacura, Casilla 19001, Santiago 19, Chile\label{inst2} 
\and 
Dpto. Astrof\'{\i}sica, Centro de Astrobiolog\'{\i}a (INTA-CSIC), ESAC Campus, PO Box 78, 28691, Villanueva de la Ca\~nada, Spain \label{inst3} 
\and 
Max Planck Institut f\"ur Astronomie, K\"onigstuhl 17, 69117, Heidelberg, Germany \label{inst4} 
\and 
Departamento de F\'isica y Astronom\'ia, Facultad de Ciencias, Universidad de Valpara\'iso, Av. Gran Breta\~na 1111, 5030 Casilla, Valpara\'iso, Chile\label{inst5} 
\and
Centro de Radioastronom\'ia y Astrof\'isica, Universidad Nacional Aut\'onoma de M\'exico, P.O. Box 3-72, 58090 Morelia, Michoac\'an, M\'exico\label{inst6} 
\and 
Academia Sinica, Institute of Astronomy and Astrophysics, P.O. Box 23-141, Taipei  106, Taiwan \label{inst7}
\and
Herschel Science Centre, ESAC-ESA, PO Box 78, 28691 Villanueva de la Ca\~nada, Madrid, Spain \label{inst8}
\and
Depto. F\'{\i}sica Te\'orica,  Facultad de Ciencias, Universidad Aut\'onoma de Madrid,  E-28049 Madrid, Spain \label{inst9}
}

   \date{Accepted por publication in A\&A}

  \abstract
  % {}  leave it empty if necessary  
   {Chamaeleon II molecular cloud is an active star forming region that offers an excellent opportunity for studying the formation of brown dwarfs in the southern hemisphere.}
   {Our aims are to identify a population of pre- and proto- brown dwarfs (5$\sigma$ mass limit threshold of $\sim$0.015~M$_{\odot}$) and provide information on the formation mechanisms of substellar objects.}
   {We performed high sensitivity observations at 870~$\mu$m using the LABOCA bolometer at the APEX telescope towards an active star forming region in Chamaeleon II. The data are complemented with an extensive multiwavelength catalogue of sources from the optical to the far-infrared to study the nature of the LABOCA detections.}
   {We detect fifteen cores at 870~$\mu$m, and eleven of them show masses in the substellar regime. The most intense objects in the surveyed field correspond to the submillimeter counterparts of the well known young stellar objects DK Cha and IRAS 12500-7658.  We identify a possible proto-brown dwarf candidate (ChaII-APEX-L) with IRAC emission at 3.6 and 4.5 ~$\mu$m.}
   {Our analysis indicates that most of the spatially resolved cores are transient,  and that the point-like starless cores in the sub-stellar regime (with masses between 0.016~M$_{\odot}$ and 0.066~M$_{\odot}$)  could be pre-brown dwarfs cores gravitationally unstable if they have radii smaller than 220 AU to 907 AU (1.2$"$ to 5$"$ at 178 pc) respectively for different masses.   ALMA observations will be the key to reveal the energetic state of these pre-brown dwarfs candidates.}

\authorrunning{de Gregorio-Monsalvo et al.}
\titlerunning{A search for pre- and proto-brown dwarfs in Chamaeleon II} 
\maketitle
%
%________________________________________________________________

\section{Introduction}
The formation of brown dwarfs (BDs) is still a hot topic of research.  The most widely discussed scenarios for their formation include turbulent fragmentation \citep{Pad04}, ejection from multiple protostellar systems \citep{Rei01,Bat02,Bat12}, disk fragmentation and subsequent ejection \citep{Sta09}, and photo-evaporation of massive pre-stellar cores \citep{Whi04}. 
Since stars and brown dwarfs evolve very rapidly during the first million years, many answers to the formation mechanism (or mechanisms) must come from the study of their properties when they are deeply embedded in the natal cloud, what we call the 'proto-BD' stage (which would correspond to the Class 0/I stage in the classical evolutionary scheme of young stellar objects; \citealt{Lad87,And93,And00}).  If we find proto-BDs surrounded by substantial envelopes, similar to the ones observed in the first stages of low-mass protostars \citep{And00}, they would provide direct support for in situ formation of brown dwarfs.  Thus, sensitive millimeter and submillimeter surveys offer the exciting prospect of identifying a population of proto-BD candidates for the first time. 

Previous works aiming to study the earliest stages of the formation of  BDs in the millimeter/submillimeter regime have followed two different approaches: core surveys (map the dust continuum emission of a region of a molecular cloud to detect the presence of sub-stellar cores, limited by the signal-to-noise ratio achieved and the extension of the maps; e.g. \citealt{Gre03}), and the study of individual objects using previous information at other wavelengths (e.g., \citealt{Bar09,And12,Pal12,Lee13,Pal14}). 

\begin{figure*}[ht!]
\centering
     \includegraphics[width=8cm, angle=-90]{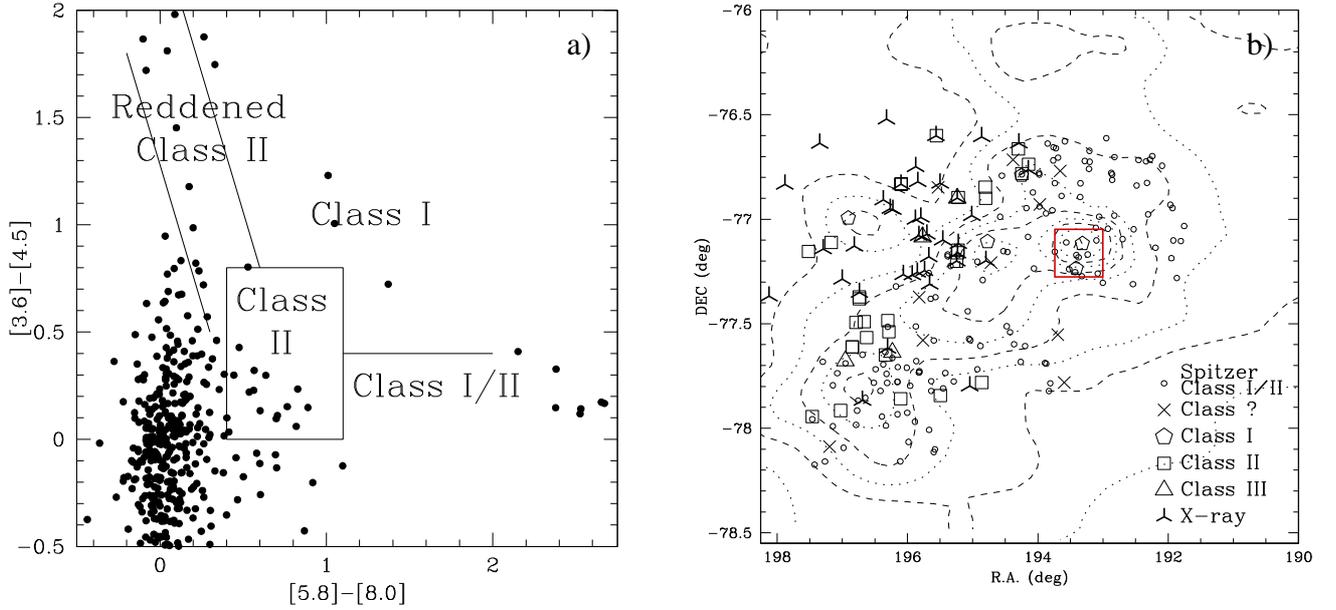}
     \caption{a) Spitzer/IRAC Color-Color diagram, with a classification for Class I, II, and III for candidate members in Cha~II. b) Distribution of possible members of Cha~II on the sky. The contours correspond to an IRAS map at 100 microns. X-ray sources appear as three-point stars (mostly Weak-line TTauri stars), from the literature. New candidate members, based on our analysis of Spitzer data, are displayed as triangles, squares, pentagons, crosses and small circles (sources with evolutionary stages between Class I and II), depending on the evolutionary class (i.e., the IRAC IR excess, from a Color-Color diagram following \citealt{All04}). Red square shows the region selected to be observed with LABOCA, which includes two out of the three Class I objects of the Cha II cloud.}
     \label{region_selection}
\end{figure*}

The development of the new generation of submillimeter bolometer arrays like LABOCA\footnote{http://www.apex-telescope.org/bolometer/laboca/} at the APEX\footnote{http://www.apex-telescope.org/} telescope together with the excellent weather conditions of the site  (Llano Chajnantor, 5000~m high), provide an excellent opportunity to carry out core surveys mapping extended regions at high sensitivity in a reasonable amount of time. 

The southern star forming region in Chamaeleon consists of three major dark clouds --\object{Chamaeleon} I, II \& III-- and a handful of smaller structures \citep{Sch77}.  Given its relative proximity (160--180 pc), youth (1--3 Myr) and intermediate galactic latitude ($b \sim -17\deg$), this cloud complex is well suited for investigations of low-mass stellar and sub-stellar populations.

We are particularly interested in \object{Chamaeleon II} (\object{Cha II}) region since it seems to be in the earliest stage of evolution in comparison with \object{Cha I} and \object{Cha III} regions, which offers an excellent opportunity for studying the formation of brown dwarfs:  a number of pre-main-sequence stars
with H$\alpha$ emission were identified in this region \citep{Sch77,Har93}. Near-infrared and IRAS data confirmed that these sources likely harbor circumstellar material (e.g. \citealt{Lar98}).

An ISO\footnote{http://iso.esac.esa.int/} study revealed four new candidate young stellar objects, including a very low luminosity ($\ge$ 0.01 $L_\odot$) source with mid-infrared excess \citep{Per03}. \cite{Alc00} detected 40 X-ray sources using ROSAT\footnote{http://science.nasa.gov/missions/rosat/} observations (only 14 of these sources coincided with previously known young stellar objects) and suggested that Cha~II contains a smaller number of weak-line T Tauri stars (Class III) than classical T Tauri stars (Class II), unlike the rest of Chamaeleon major dark clouds (Cha~I and III), suggesting that Cha~II is the one in the earliest stage of evolution. In addition, \cite{Vuo01} using $I, J, K_s$ data from the DEep Near Infrared Survey of the southern sky (DENIS\footnote{http://irsa.ipac.caltech.edu/Missions/denis.html}; \citealt{Epc97}) identified 51 candidate low-mass stars and brown dwarfs in the region. Comparison with theoretical evolutionary tracks suggests that their photometrically-selected candidates have masses $<$ 0.2 $M_\odot$ and ages 1--10 Myr.  \cite{Bar04} collected spectra for some of them and confirmed the presence of a classical T Tauri object with signatures of disk accretion and outflow near the substellar domain.  Cha~II has been surveyed during the course of the Spitzer Legacy program “From Molecular Cores to Planet-forming Disk“ (c2d; \citealt{Eva03}) and as part of the Herschel Gould Belt survey key project \citep{And10,Spe13}. 

In this work we present very deep observations at 870~$\mu$m using the LABOCA bolometer at the APEX telescope  towards the Cha II complex. The sensitivity of our map ($\sim$5 mJy/27.6$''$ beam) provides a 5$\sigma$ threshold for the H$_{2}$ column density of 2.1$\times$10$^{21}$ cm$^{-2}$, and a 5$\sigma$ mass limit of $\sim$0.015 M$_{\odot}$, well within the substellar regime. This is the first time that such a deep map has been done in this region.  Our goal is to detect faint cold dust envelopes surrounding very low mass objects to provide the best candidates to pre-brown dwarfs (the scaled-down version of the low mass pre-stellar cores) and proto-brown dwarfs in the Cha~II region. We have complemented our submillimeter data with a multiwavelength database, from the optical to the far-infrared, aiming to reveal the true nature of the detections and to provide clues for their properties and formation mechanism.

This paper is structured as follows: in Section~\ref{Observations}, we describe the observations and the archival data. In Section~\ref{Results}, we show the results. Finally, we present the discussion and the conclusions of this work in Sections~\ref{Discussion} and ~\ref{Conclusions}.

\section{Observations and Archival data}
\label{Observations}

\subsection{Selection of the region}
The region in the cloud with most recent star formation was selected to carry out our study. Using  InfraRed Array Camera (IRAC) Spitzer archival data we generated a color-color diagram (see Fig~\ref{region_selection}, left panel) that was used to classify the stage of the evolution of Cha II members in Class I, II and III sources, following \citet{All04}. Then, we represented the spatial distribution of the known candidate young members of Cha~II, and  we selected a region of $14'\times14'$ projected size where the highest density of very low mass ClassI/II and Class I objects was observed  (see Fig~\ref{region_selection}, right panel).  Two out of the three Class I sources in the area, according to their positions in the color-color diagram, lie in the region of our LABOCA observations.

\begin{table*}[ht!]
\small
\caption{\label{multiwave} Archival data used to perform a multiwavelegth study in the cores detected at 870~$\mu$m}
\centering
\begin{tabular}{lccccccccc}
\hline\hline
Telescope/& Filter$_{1}$ ($\lambda$$_{1}$) & Filter$_{2}$ ($\lambda$$_{2}$) & Filter$_{3}$ ($\lambda$$_{3}$) & Filter$_{4}$ ($\lambda$$_{4}$) & Filter$_{5}$ ($\lambda$$_{5}$)   \\

Instrument or survey  & & & & &  \\

\hline
\hline

ESO 2.2m/WFI\tablefootmark{a} &Rc (651.7 nm)      &H$\alpha$ (658.8 nm)   &NB665 (665.6 nm)        &SII (676.3 nm)   &Ic (783.8 nm)   \\
ESO 2.2m/WFI\tablefootmark{a} &I (826.9 nm)         &MB856 (856.2 nm)         &MB914 (914.8 nm)        &Z (964.8 nm)   \\
VLT/VIMOS\tablefootmark{a}     &I (817.1 nm)          & & & &  \\
ESO 1m/DENIS                           &I (0.8 ~$\mu$m)     &J (1.2 ~$\mu$m)              &Ks (2.1 ~$\mu$m)          & & \\
Mt. Hopkins-CTIO/2MASS         &J (1.2~~$\mu$m)    &H (1.7 ~$\mu$m)            &K (2.2 ~$\mu$m))           & & \\
WISE                                           &W1 (3.4 ~$\mu$m) &W2 (4.6 ~$\mu$m)          &W3 (12 ~$\mu$m)          &W4 (22 ~$\mu$m)  & \\
Spitzer/IRAC\tablefootmark{b}  &I1 (3.6~$\mu$m)    &I2 (4.5 ~$\mu$m)            &I3 (5.8 ~$\mu$m)           &I4 (8.0 ~$\mu$m)  & \\
Spitzer/MIPS\tablefootmark{b}  &M1 (24 ~$\mu$m)   &M2 (70 ~$\mu$m)           & & & \\
Herschel/PACS\tablefootmark{c}                        & Blue (70 ~$\mu$m)  & Red (160 ~$\mu$m)      & & & \\
Herschel/SPIRE\tablefootmark{c}                         & Blue (250 ~$\mu$m)    & Green (350 ~$\mu$m)  & Red (500 ~$\mu$m)  & & \\
Akari/IRC                                   &S9W (9 ~$\mu$m)   &L18W (18 ~$\mu$m)       & &  & \\
Akari/FIS                                    &N60 (65 ~$\mu$m) &WIDE-S (90 ~$\mu$m)   &WIDE-L (140 ~$\mu$m)   &N160 (160 ~$\mu$m)   & \\

\hline
\hline                                                        
\end{tabular}                                                 
\tablefoot{                                                   
%Bla, bla\\
\tablefoottext{a}{Archival ESO Projects 2064.I-0559, 67.C-0225(A), 077.C-0339(A), 075.C-0294(C), and 68.C-0311(A), for WFI data, and 072.C-0046(B) for VIMOS.}
\tablefoottext{b}{These data are part of the legacy project  "cores to disks"  (c2d)  (see \citealt{Eva03})}
\tablefoottext{c}{Data part of the Herschel Guaranteed Time Key Programme ``Gould Belt survey'' (KPGT\_andre\_1; PI: Philippe Andr\'e)}
}
\end{table*}

\subsection {LABOCA 870~$\mu$m continuum data}
We carried out continuum observations at 870~$\mu$m in the selected region using LABOCA bolometer array, installed on the Atacama Pathfinder EXperiment (APEX\footnote{This work is partially based on observations with the APEX telescope. APEX  is a collaboration between the Max-Plank-Institute fur Radioastronomie, the European Southern Observatory, and the Onsala Space Observatory}) telescope. 
Our data were acquired on 2010 April 08, 09, and 11 during the Onsala Swedish program O-085.F-9301A-2010 under good weather conditions (pwv$\sim$1.20 mm, equivalent to a zenith opacity of 0.34 at 870~$\mu$m). Observations were performed using a spiral raster mapping centered at $\alpha$ = 12$^{h}$53$^{m}$29.9$^{s}$, $\delta$ = -77$^{\rm o}$10$'$57$''$(J2000.0) and covering an area of projected size $14'\times14'$.  The total integration time was $\sim$13 hours (10.4 hours on source). Calibration was performed using observations of Mars as well as the secondary calibrator G305.81-0.25. The absolute flux calibration uncertainty was $\sim$8$\%$.  The telescope pointing was checked every hour and focus settings were checked once per night and during the sunset.    

Data were reduced using CRUSH\footnote{http://www.submm.caltech.edu/~sharc/crush/} software package (see \citealt{Kovacs08}) and the BOlometer Array Analysis Software (BoA\footnote{http://www.apex-telescope.org/bolometer/laboca/boa/}). The pre-processing steps consisted of flagging dead or cross-talking channels, frames with too high telescope accelerations and with unsuitable mapping speed, as well as temperature drift correction using two blind bolometers.  Data reduction process included flat-fielding, opacity correction, calibration, correlated noise removal (atmospherics fluctuations seen by the whole array, as well as electronic noise originated in groups of detector channels), and de-spiking.  Every scan was visually inspected to identify and discard corrupted data. 

Data processing was optimized to recover faint sources. The final map was smoothed to a final angular resolution of $27.6''$. As a result, we obtained a 1$\sigma$ point source sensitivity of $\simeq$4 mJy at the center of the map and $\simeq$6 mJy at the edges.

\subsection {Mining the archives: optical to far-infrared data}

In order to properly classify the sources detected at 870~$\mu$m  and search for optical to far-infrared counterparts, we built an extensive multiwavelength catalogue in the Cha II region with public data from different space missions and ground-based telescopes.  We used {\it optical} data from the Wide Field Imager (WFI\footnote{http://www.eso.org/sci/facilities/lasilla/instruments/wfi.html}) at MPG/ESO 2.2m telescope, as well as the Visible Multi-Object Spectrograph (VIMOS\footnote{http://www.eso.org/sci/facilities/paranal/instruments/vimos.html}) in the imaging mode, at the VLT. 
In the {\it near-infrared}, we used the point-source catalogue from the 2 Microns All sky survey (2MASS\footnote{http://irsa.ipac.caltech.edu/Missions/2mass.html}), and  the Deep Near Infrared Survey of the Southern Sky (DENIS).  The Preliminary Release Image and Source data product from the Wide-field Infrared Survey Explorer (WISE\footnote{\url {http://www.nasa.gov/mission_pages/WISE/main/index.html}}) mission,  Spitzer IRAC\footnote{http://www.spitzer.caltech.edu/mission/398-The-Infrared-Array-Camera-IRAC-} data  and Akari\footnote{http://irsa.ipac.caltech.edu/Missions/akari.html} Infrared Camera (IRC) Point Source Catalogue data products were considered in the {\it mid-infrared} range. Data in the {\it far-infrared}  were taken from the Multiband Imaging Photometer for Spitzer  (MIPS\footnote{http://www.spitzer.caltech.edu/mission/396-The-Multiband-Imaging-Photometer-MIPS-}) and from the Akari Far-Infrared Surveyor (FIR) Bright Source Catalogue  (see Table~\ref{multiwave} for a detailed description).  Herschel PACS\footnote{http://www.mpe.mpg.de/ir/Pacs} and SPIRE\footnote{http://www.astro.cardiff.ac.uk/research/astro/instr/projects/?page=spire} maps were also processed for checking the spatial coincidence of the cores detected with LABOCA in the filamentary and clumpy structure observed in Herschel maps.

\begin{figure*}[!ht]
\centering
     \includegraphics[width=19cm]{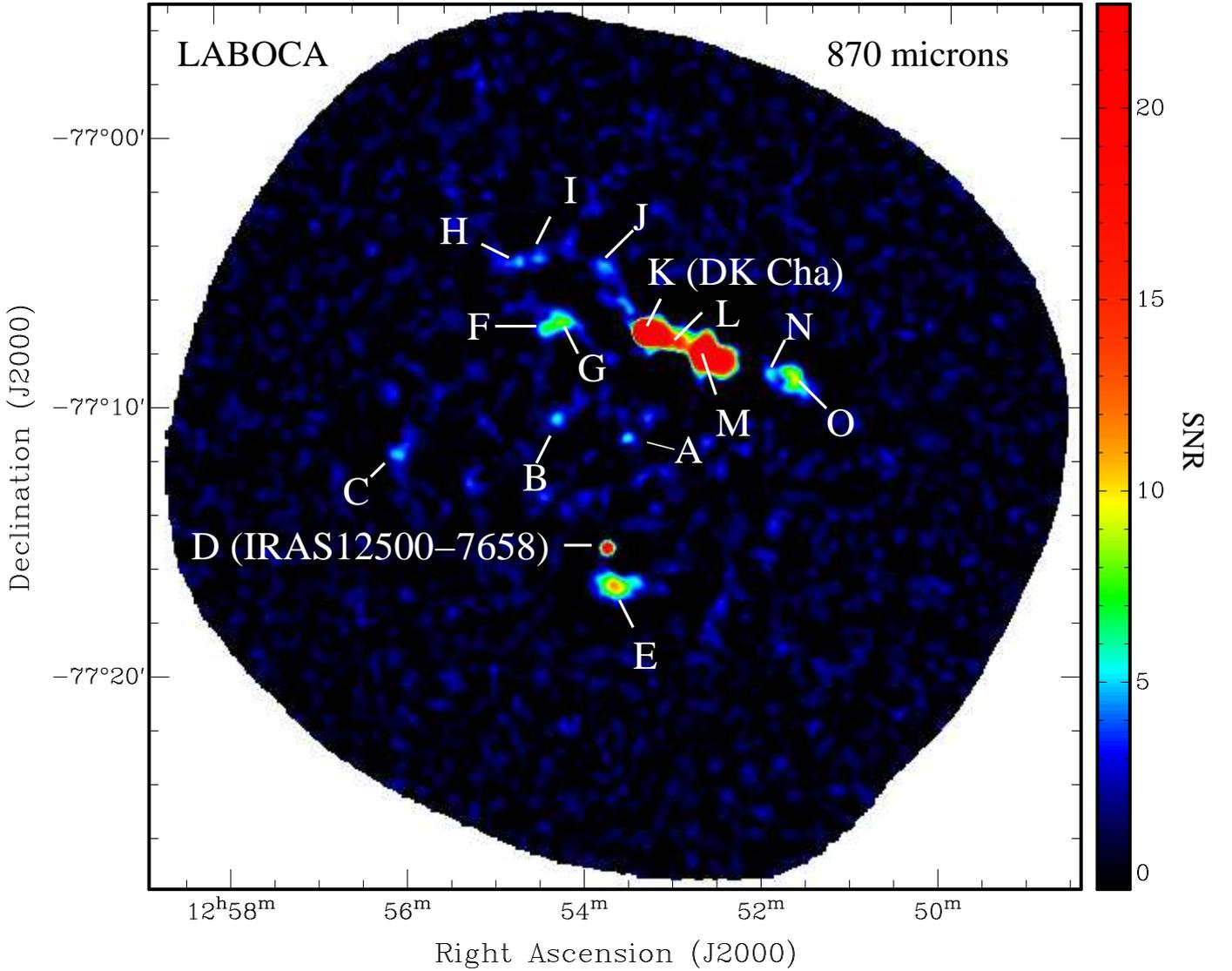}
     \caption{Signal-to-noise ratio emission map at 870~$\mu$m observed with LABOCA in the selected area of Cha II region.} 
     \label{laboca}
\end{figure*}

% Note: On-line figure 
\begin{figure*}
\centering
     \includegraphics[width=18cm]{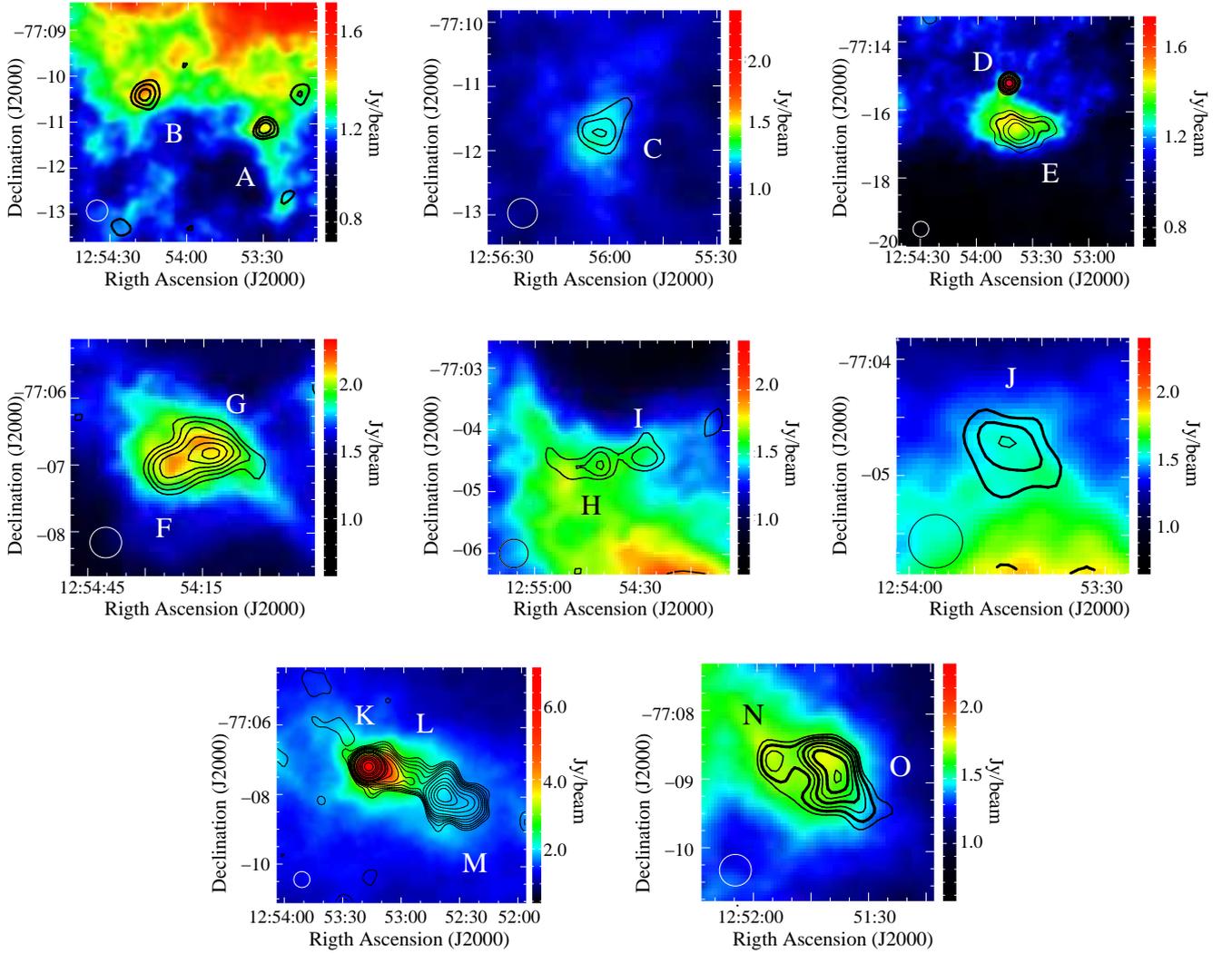}
     \caption{Close-up with details of the submillimeter cores with a SNR $\geq$5.  Colors represent the Herschel PACS map at 160 ~$\mu$m superposed in the background. Contours represent LABOCA emission at 870~$\mu$m. Lower contour represents 3$\sigma$ emission, with an increment of 1$\sigma$ between contours for all cores, except for the one corresponding to cores K (DK Cha), L, and M, with contours at 3$\sigma$, 5$\sigma$, 7$\sigma$, 9$\sigma$, 12$\sigma$, 15$\sigma$, 20$\sigma$, 30$\sigma$, 40$\sigma$, 50$\sigma$, 60$\sigma$, 90$\sigma$, 120$\sigma$, 170$\sigma$, 220$\sigma$, 270$\sigma$, 350$\sigma$, and cores  D (IRAS 12500-7658) and E, with contours at 3$\sigma$, 5$\sigma$, 7$\sigma$, 9$\sigma$, 12$\sigma$, 15$\sigma$, and 20$\sigma$. See different values of rms in table~\ref{laboca-clumps}. }        
     \label{labocaII}
\end{figure*}

\begin{figure*}
\centering
     \includegraphics[width=15.2cm]{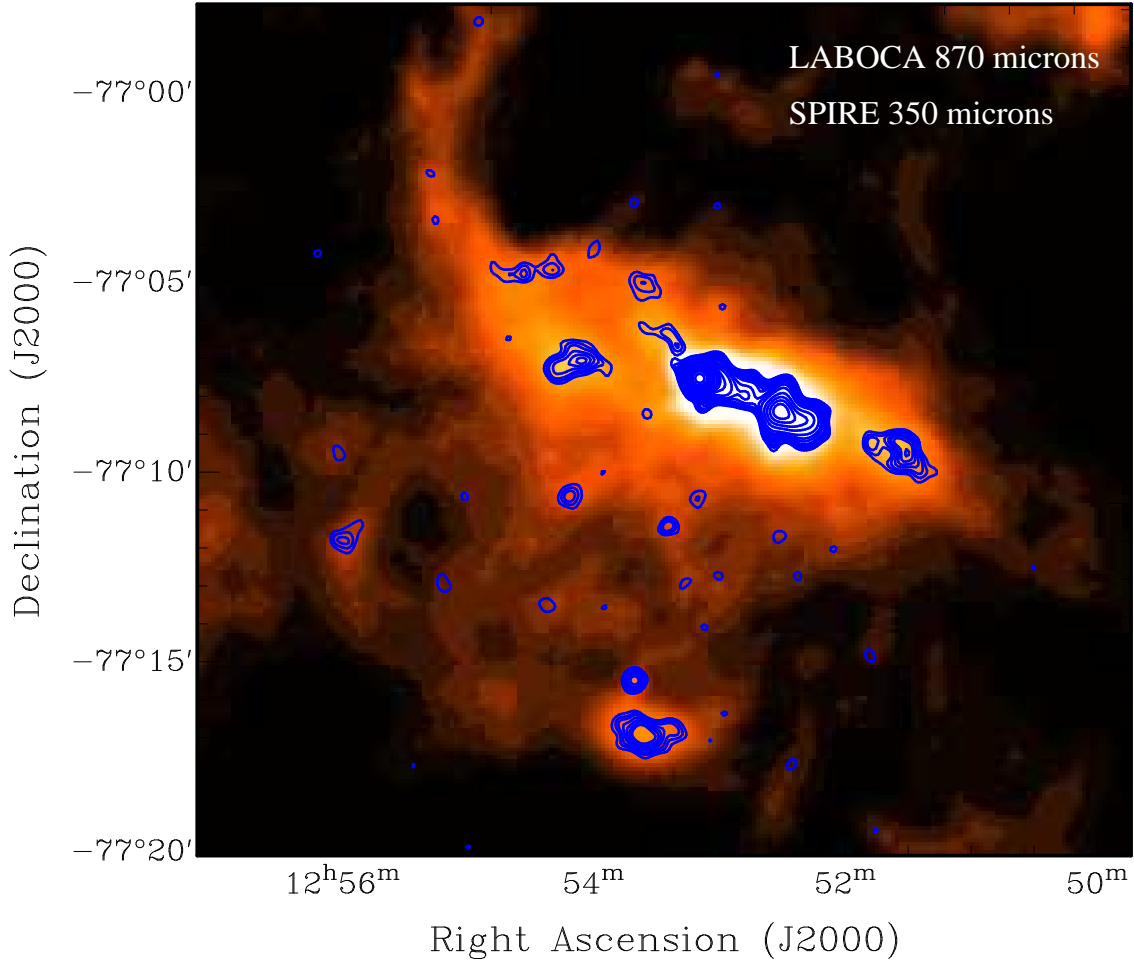}
     \caption{ SPIRE Herschel emission map at 350~$\mu$m (colour) superposed with LABOCA emission at 870~$\mu$m (contours; levels similar to DK Cha panel in Fig~\ref{laboca}).}               
     \label{herschel}
\end{figure*}

\begin{table*}
\caption{\label{laboca-clumps} Properties of the submillimeter cores detected at 870~$\mu$m in Cha II.}
\scriptsize
\centering
\begin{tabular}{ccccccccccc}
\hline\hline
Core&R.A.\tablefootmark{a} &Dec.\tablefootmark{a} &S$_{870_{\mu m}}$\tablefootmark{b}& rms &Mass & Deconvolved \tablefootmark{c} &n$_{core}$   &n$_{crit}$\tablefootmark{d} & Sep.\tablefootmark{e}  & Counterpart?\\
  &(J2000)      &(J2000)      &    (mJy)   &(mJy)          & ($M_{\sun}$)   &ang. size   & (cm$^{-3}$)  & (cm$^{-3}$)  & (``) &\\        
\hline
 
\object{ChaII-APEX-A}       &12:53:30.5    &-77:11:07    &  26     &5     &0.016      &Point-like                              &...   &4.3$\times$10$^{7}$   & 7$"$,10$"$ & No (Background star)\\
%%ChaII-APEX-2     &12:53:16.9    &-77:10:23    & 19       &4.5  &0.0115      & Point-like                            &...   &     &12$"$,17$"$ & Background star, ?\tablefootmark{d}\\
\object{ChaII-APEX-B}       &12:54:17.3    &-77:10:23    & 26       &5     &0.016      & Point-like                              &...     &4.3$\times$10$^{7}$    &15$"$ &  No (Background star)  \\
%ChaII-APEX-4     &12:54:26.0    &-77:13:19    & 19       &4.8  &0.0115      & Point-like                             &...     &    &15$"$ &Background star\\
%ChaII-APEX-5    &12:55:15.3    &-77:12:47     &  27      &5.2  &0.016.      & 35$"$                                     &      &    &15$"$  & I2$>$I1\\
\object{ChaII-APEX-C}     &12:56:03.3    &-77:11:43     &  36\tablefoottext{f}      &6     &0.022      & 35$"$           & 2.2$\times$10$^{4}$        &2.3$\times$10$^{7}$   &12$"$, 14$"$ & No (Background star) \\
\object{ChaII-APEX-D}      &12:53:42.6    &-77:15:11     &  109   &5     &0.066      &Point-like                              &...      &2.5$\times$10$^{6}$    &1$"$  & Yes (YSO IRAS 12500-7658) \\
\object{ChaII-APEX-E}     &12:53:39.0    &-77:16:35     &  460 \tablefoottext{f}    &5     &0.279    &75$"$              & 2.8$\times$10$^{4}$              &1.4$\times$10$^{5}$    &... & No  \\
\object{ChaII-APEX-F}      &12:54:21.9    &-77:06:59     &  35      &5     &0.021      & Point-like                              &...     &2.5$\times$10$^{7}$    &... & No \\      
\object{ChaII-APEX-G}    &12:54:12.3    &-77:06:47     &  130\tablefoottext{f}     &5      &0.079      & 45$"$           & 3.7$\times$10$^{4}$         &1.7$\times$10$^{6}$    &11$"$  & No (Background star)\\
\object{ChaII-APEX-H}      &12:54:40.8    &-77:04:35     &  49\tablefoottext{f}     &6      &0.030      & 35$"$          & 3.1$\times$10$^{4}$     &1.2$\times$10$^{7}$    &4$"$,14$"$,16$"$ & No (Background star) \\
\object{ChaII-APEX-I}       &12:54:27.7    &-77:04:27     &  30      &6     &0.018      & Point-like                              &...          &3.4$\times$10$^{7}$    &13$"$ & No (Background star) \\
\object{ChaII-APEX-J}       &12:53:44.8    &-77:04:43     &  39\tablefoottext{f}       &6     &0.024      &40$"$            &1.5$\times$10$^{4}$             &1.9$\times$10$^{7}$    &18$"$ & No (Background star)\\
%ChaII-APEX-14  &12:53:31.7    &-77:06:03     &  24      &5.1   &14.6      &Point-like                                   &...           &   &14$"$ &?\tablefootmark{d} \\
%ChaII-APEX-15  &12:53:28.1    &-77:06:19     &  25      &5.1   &15.2      &Point-like                                  &...            &    &... &... \\
\object{ChaII-APEX-K}      &12:53:17.3    &-77:07:11     &1745    &4     &1.059  & Point-like                                  &...            &9.7$\times$10$^{3}$    &0.5$"$  & Yes (YSO DK Cha) \\
\object{ChaII-APEX-L}      &12:52:55.8    &-77:07:35     & 78       &5     &0.047      & Point-like                              &...             &4.9$\times$10$^{6}$    &1$"$ & Yes (Proto-BD candidate) \\
\object{ChaII-APEX-M}    &12:52:39.0    &-77:07:59     & 3700\tablefoottext{f}    &4     &2.246  & 80$"$               & 1.8$\times$10$^{5}$                  & 2.2$\times$10$^{3}$   &... & No\\
\object{ChaII-APEX-N}     &12:51:55.8    &-77:08:43     & 30       &5     &0.018      & Point-like                             &...                   &3.4$\times$10$^{7}$     &16$"$ & No (Background star) \\ 
%ChaII-APEX-20  &12:51:42.6    &-77:08:39     & 49       &5.6  &29.7      &Point-like                                 &...                   &     &5$"$,17$"$ &Background stars \\
\object{ChaII-APEX-O}    &12:51:39.0    &-77:08:55     & 260\tablefoottext{f}       &6    &0.158      &55$"$           &4.0$\times$10$^{4}$        & 4.4$\times$10$^{5}$    &13$"$, 16$"$ & No (Background star) \\

\hline
\hline                                                        
\end{tabular}                                                 
\tablefoot{ 
\tablefoottext{a} {Position of the maximum emission of the cores. Units of right ascension are hours, minutes, and  seconds. Units of declination are degrees, arcminutes, and arcseconds.} 
\tablefoottext{b}{Flux density computed above 3$\sigma$ emission.}
\tablefoottext{c}{Deconvolved size after fitting a Gaussian to the cores. The averaged of the major and minor axis is represented.}
\tablefoottext{d}{Critical density of the core, calculated following the equation in Section~\ref{pre-BD}.}
\tablefoottext{e}{Separation of the counterpart candidates to the nominal position of the submillimeter core.} 
\tablefoottext{f} {For cores showing extended emission we applied a specific data processing (using option "extended" in CrUSH) for avoiding the filter of extended structures.}
}

\end{table*}

\section{Results}
\label{Results}
\subsection {Submillimeter emission at 870~$\mu$m}

The LABOCA map reveals a clumpy distribution of cold dust with two bright sources spatially coincident with well known young stellar objects (YSOs; see Fig.~\ref{laboca}). The northern one, known as \object{DK Cha}, is associated with \object{IRAS 12496-7650} and  the southern one is the submillimeter counterpart of \object{IRAS 12500-7658}.   Both sources are surrounded by more extended diffused emission, probably tracing material from the parental cloud. 

The central part of the map, which is the region with the lowest noise, shows various faint cores.  We have identified a total of fifteen local maxima at more than 5$\sigma$ emission (see Table~\ref{laboca-clumps} for a summary of the main properties of the detections).  Eight of them correspond to unresolved point-like sources and the other seven cores show a more extended morphology.

\subsection {Searching for optical to infrared counterparts}

We used our extensive multiwavelength catalogue to search for counterparts to the detections at 870~$\mu$m (see some examples in Table~\ref{multiwave-example}). 
The position of every submillimeter source was cross-matched with this catalogue, considering all possible counterpart candidates falling into the beam size of LABOCA (27.6$"$) for point-like sources, and into the total core area for the ones showing spatial structure.  
Once we identified all the counterpart candidates as a function of the separation to the submillimeter core,  we analyzed their spectral energy distribution to identify and classify reliable candidates. An extended explanation of the used method is provided in appendix~\ref{method}. Results on this method are shown in Table~\ref{laboca-clumps}.

\subsection{Mass of the cores}

For the calculation of the mass of the cores we assume the emission at 870~$\mu$m comes totally from thermal dust and it is optically thin. We estimate the total mass (gas+dust) assuming a gas-to-dust ratio of 100 and adopting the following relation: 
$$M = \frac{S_{870}~ d^{2}}{\kappa_{870} ~ B_{870}(T_{d})}\;$$
where $S_{870}$ is the flux density of the cores (i.e. the integrated emission over the solid angle) at 870~$\mu$m, $d$ is the distance to the source, $\kappa$$_{870}$ is the dust opacity per unit mass (gas+dust) column density at 870~$\mu$m, and $B_{870}(Td)$ is the Planck function at 870~$\mu$m for a dust temperature $T_{d}$.  
We adopt a distance of 178 pc \citep{Whi97}, a $T_{d}$=10 K, and a dust opacity at 870~$\mu$m of 0.0175 cm$^{2}$ gr$^{-1}$ that results from interpolating the values at 700 and 1000~$\mu$m given in column 6 of Table 1 of \cite{Ose94} (corresponding to dust grains with thin ice mantles at a density of 10$^6$~cm$^{-3}$). The opacity law of \cite{Ose94} is consistent with the law inferred from Herschel data towards Infrared Dark Clouds (e.g., \citealt{Lim14}). The value adopted by us at 870~$\mu$m is $\lesssim$50$\%$ compared to the value adopted by the Herschel Gould Belt Survey (of $\sim$0.01189~cm$^2$\,g$^{-1}$, \citealt{Roy13}), well below the typical uncertainty for the dust opacity inferred from modeling of Herschel data, of about a factor of two \citep{Wag15}.

Flux density was calculated by integrating the emission above 3 $\sigma$ contours. 
The derived core masses are affected by errors coming mainly from the uncertainty in $T_{d}$, dust opacity, and the distance.  The error in the distance calculation is estimated to be 18 pc  \citep{Whi97}, providing a change in the core masses estimation of a factor $\sim$1.2. An increment of a factor two in the value of the opacity and of the $T_{d}$ produces a decrement in the value of the mass of a factor $\sim$2 and $\sim$3.3 respectively.  Therefore the highest source of error for the estimation of masses is the value of the dust temperature.   Previous works aiming to model the dust temperature profiles of pre-stellar cores provided a variation from 13 K in the outer parts and 7 K in the inner parts (see \citealt{Eva01}). In this work  we adopt an average value of $T_{d}$ =10 K. The adopted criteria provides eleven cores with masses in the substellar regime and four cores with stellar masses (see Table~\ref{laboca-clumps}).

\section{Discussion}
\label{Discussion}

\subsection{Nature of the counterpart candidates}
In order to reveal the nature of the optical to infrared possible counterparts found, we performed a study of the SED properties on every reliable counterpart candidates of the submillimeter cores using \citet{Rob06,Rob07} SED tool. 
From this analysis we derive that two cores (D and K) are clearly associated with the well known young objects DK Cha (IRAS 12496-7650) and IRAS 12500-7658. Nine cores (A, B, C, G, H, I, J, N, and O)  lack any YSO candidate counterpart, being the SEDs of the found possible counterparts compatible with extincted background stars and hence not associated with the submillimeter cores.  Cores E, F, and M do not show any counterpart.  In core L, we found a YSO candidate with counterparts in I1 and I2 IRAC bands.  These results are summarized in Table~\ref{laboca-clumps}.

The NASA/IPAC Extragalactic Database (NED\footnote{https://ned.ipac.caltech.edu/}) was checked to search for extragalactic contamination at the position of each core, and no extragalactic counterparts were found.   We also compared our map with recent public Herschel observations with SPIRE and PACS, obtaining a very good spatial coincidence between the cores at 870~$\mu$m and the clumpy filament emission shown by Herschel (see Fig.~\ref{labocaII} for a close up superposition of every core detected at 870~$\mu$m with PACS at 160~$\mu$m and Fig.~\ref{herschel} for an overlap of the SPIRE map at 350~$\mu$m and the region observed with LABOCA). It is remarkable the faint dusty micro-filaments where most of the LABOCA cores seem to be embedded. These filaments, observed at a noise level in the LABOCA map and clearly confirmed by Herschel, suggest a scaled-down version of the network of longer and dense filaments recently studied by \citet{And10}, \citet{Kon10} and \citet{Hac13} in several molecular clouds, and proposed to be the precursors of pre-stellar cores via fragmentation.

\subsection{Previously known young stellar objects in the region}
\begin{figure*}[ht!]
\centering
     \includegraphics[width=17cm, angle=0]{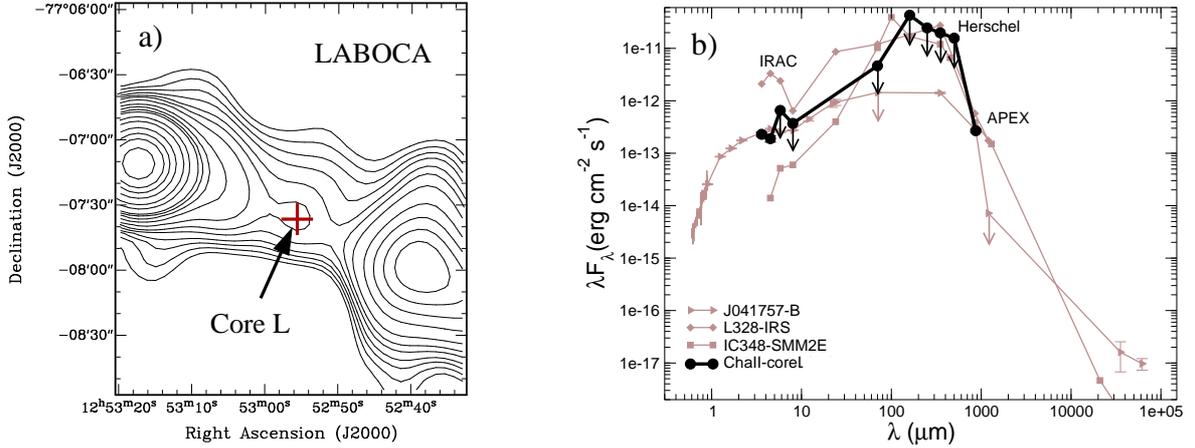}
     \caption{a) Close-up of LABOCA emission at core L (similar contour levels as in Fig~\ref{laboca}). Red cross represents the position of the IRAC 2 Spitzer emission at the position of core L. b) Spectral energy distribution of core L (black thick line) and the proto-BD candidates J041757-B, L328-IRS, and IC348-SMM2E (brown thin lines; \citealt{Pal12}, \citeyear{Pal14}; \citealt{Lee13}). Arrows represent upper limits. }
     \label{SEDL}
\end{figure*}

\indent IRAS 12496-7650, also known as DK Cha, is the most massive object in Cha II region, an intermediate-mass Herbig Ae star in a transitional stage of evolution between  Class I and II \citep{Spe08} that hosts a face-on disk-outflow system \citep{Van10}. This source has been extensively studied by \cite{Spe13}, providing an estimation of the most probable parameters of the disk surrounding DK Cha. \cite{Spe13} model the SED of this source from the optical to millimeter wavelengths (using Herschel data) and the flux at 870 $\mu$m expected from their set of best-fitting models agrees very well with our measurements given in Table~\ref{laboca-clumps}.

Our deep submillimeter observations reveal this source immersed in a well defined elongated and clumpy nebulosity. Elongated structures with multiple sub-peaks have been observed surrounding other intermediate-mass YSOs, being the nature of this emission not clear whether  due to multiple sources embedded in the elongated structure, or due to shocks in the outflow, or just tracing the natal cloud (e.g. \citealt{Fue07,Rat05}). In the case of DK Cha the outflow has been proposed to be perpendicular to the plane of the sky, which rules out this phenomenon as cause of the dust structure observed.   Considering the filamentary structures observed in Herschel maps with multiple cores embedded inside,  such a stream of dust clumps may be tracing embedded YSOs being formed inside the natal cloud.  

Besides DK Cha there is another young central object in the elongated dust structure, which corresponds to the infrared counterpart of core L (see subsection~\ref{proto}). A high level of fragmentation is revealed in our map and one of the embedded cores (core M) seems to be gravitationally unstable (see subsection~\ref{pre-BD}). Therefore we conclude they are multiple pre-stellar and protostellar cores belonging to the same parental nebula. 

IRAS 12500-7658 has been catalogued as a Class I object of spectral type close to K5  \citep{Spe13}. In this work, we detect compact submillimeter emission associated with this IRAS source very close to a southern and extended starless core of $\simeq$10$^{4}$ AU size (see Fig.~\ref{laboca} and ~\ref{herschel}), typical scale of dense cores where low mass stars are formed.   \cite{Spe13} modeled in detailed its SED and inferred a first estimation of the parameters of the young disk surrounding this source, predicting a flux at 870 micron consistent with our observed value (Table~\ref{laboca-clumps}). No further analysis will be presented in this work beyond the SED and the photometry shown in Fig~\ref{counterparts} and Table~\ref{multiwave-example}.

\subsection{A possible candidate to proto-BD}
\label{proto}

The infrared counterpart (I1, I2) of core L lies only about $\sim1''$ away from the peak of the submillimeter source. This excellent match, which can be seen in Fig~\ref{SEDL}-left, strongly suggests that the infrared and submillimeter sources are spatially associated. 
In order to further explore the possible nature of core L, we compared its SED (for which we list the photometry in Table~\ref{photometryL}) to the known SEDs of previously reported Class I (\object{J041757B}, \citealt{Bar09,Pal12}) and Class 0 \citep{Lee13,Pal14} proto-BD candidates, and the result is shown in Fig~\ref{SEDL}-right. The figure shows that the IRAC and 870~$\mu$m fluxes are comparable to the fluxes of the Class I proto-BD candidate J041757B. But core L could still be a Class 0 proto-BD, depending on the real values of the fluxes between 5 and 500~$\mu$m, for which only upper limits could be estimated, due to the close position of DK Cha dominating the emission at these wavelengths. The upper limit to the bolometric luminosity of core L is $<0.08$~$L_\mathrm{\odot}$, similar to the bolometric luminosity of the Class 0 proto-BD \object{IC348-SMM2E} \citep{Pal14}. Such a low bolometric luminosity, together with the fact that its envelope mass is substellar ($\sim50$~$M_\mathrm{Jup}$) indicates that core L could be a proto-BD candidate. We warn that this is still a speculative result based on three photometry points (in the near infrared and at submillimeter frequencies) and a set of upper limits,  but due to the scarce number of known proto-BD candidates the core ChaII-APEX-L deserves further studies to confirm its real nature.  

\begin{table}
\caption{\label{photometryL} Photometry for core L}
%\small
\centering
\begin{tabular}{cccl}
\hline\hline
$\lambda$    &S$_{\nu}$     &$\sigma$    &Instrument  \\
(~$\mu$m)     & (mJy)          & (mJy)          &              \\        
\hline
3.6        &0.276    & 0.025    &Spitzer/IRAC\\
4.5        &0.29    & 0.04      &Spitzer/IRAC\\
5.8        & $<$1.3            &...       &Spitzer/IRAC\\
8.0        & $<$1.0                &...       &Spitzer/IRAC\\
70         & $<$108                &...      &Herschel/PACS\\
160       & $<$2290                &...    &Herschel/PACS\\
250       & $<$2050                &...     &Herschel/SPIRE\\
350       & $<$2310               &...      &Herschel/SPIRE\\
500       & $<$2610                 &...    &Herschel/SPIRE\\
870       &78               &   4.9          &APEX/LABOCA\\
\hline
\hline                                                        
\end{tabular}                                                 
%\tablefoot{ }

\end{table}

\subsection{Cores with no counterpart: pre-BD cores?}
\label{pre-BD}

 \citet{Pad04}, who investigated the in-situ formation of BDs in supersonically turbulent clouds via turbulent fragmentation, stated that substellar objects can be formed in cores of a density as high as the critical one for the collapse of a BD mass core. These authors used the mass of the critical Bonnor-Ebert isothermal sphere to infer the critical density:

$$n_{crit} = \left(\frac{3.3}{M_{BE}}\right)^{2} \left(\frac{T}{10}\right)^{3} 10^{3},$$  with $M_{BE}$ in M$_{\odot}$, $T$ in K, and $n_{crit}$ in cm$^{-3}$.

For the resolved extended starless cores C, E, G, H, J, and O we obtain a $n_{core} < n_{crit}$ (see Table~\ref{laboca-clumps}), suggesting all of them are gravitationally stable transient cores. Core M, close to  DK Cha and with a mass in the stellar regime, is the only one that presents extended emission with a $n_{core} > n_{crit}$, with a value of n$_{core}$=1.8$~\times 10^{5}~cm^{-3}$  two orders of magnitude larger than n$_{crit}$=2.2$~\times 10^{3}~cm^{-3}$. This result suggests this core is in a pre-stellar, gravitationally unstable, stage.  

For the rest of the point-like (unresolved) sources detected in our APEX/LABOCA map,  starless (cores A, B, F, I, and N ) or associated with a central YSO (cores D,K, and L), we calculated the maximum radius required to have a core density higher than the critical one. We conclude that for objects in the sub-stellar regime radius between R$\leq$220 AU for the less massive cores (0.016 M$_{\odot}$)  and R$\leq$907 AU for the most massive one (0.066 M$_{\odot}$)  are needed to become gravitationally unstable.   

Whether these faint starless point-like sources are transient cores or will grow in mass and will become pre-stellar objects that form brown dwarfs or/and planetary-mass objects can only be answered by studies at high-angular resolution and high-sensitivity using a telescope like ALMA (a radius of 220 AU at 178 pc subtends a projected size of 1.2$''$ and ALMA can achieve a spatial resolution below 0.1$"$ at the frequencies of the study presented here). In addition spectroscopic observations similar to the ones reported by \cite{And12}  using Plateau de Bure interferometer to observe the now confirmed pre-brown dwarf Oph B-11, will provide the final answer to their energetic state and their dynamical mass.

\section{Conclusions}
\label{Conclusions}
In this work we present high sensitivity observations at 870~$\mu$m toward an active star forming region in the Chamaeleon II molecular cloud, which provides a 5$\sigma$ mass limit of 0.015 $M_\odot$, in the substellar regime. Our main goal is to identify a population of pre- and proto-brown dwarf candidates to be studied in detail with future ALMA observation.  We built an extensive multiwavelength catalogue from the optical to the far-infrared in order to search for counterparts and classify properly the sources detected at 870~$\mu$m. The main conclusion are:
\begin{itemize}
\item  A total of fifteen cores were detected in the mapped region at 870~$\mu$m, eleven of them have substellar masses and four of them show masses in the  stellar regime. 
\item Two cores are the submillimeter counterparts of the well known YSOs DK Cha and IRAS12500-7658.  We identified a possible candidate to proto-BD (core ChaII-APEX-L) that shows counterparts at 3.6 and 4.5 ~$\mu$m. The rest of the cores have a starless nature, seven of them show extended emission and five are spatially unresolved point-like sources. 
\item Based on the \cite{Pad04} theory of formation of BDs via turbulent fragmentation,  we infer that most of the spatially resolved cores are transient and gravitationally stable except core ChaII-APEX-M, which density is higher than the critical one.  For point-like starless cores in the substellar regime (with masses between 0.016 $M_\odot$ and 0.066 $M_\odot$) we conclude that radius smaller than 220 AU to 907 AU respectively are needed to be in a gravitationally unstable stage and became pre-brown dwarfs. 
\end{itemize}
  
This paper, based on very deep observations at 870~$\mu$m, provides a sensitivity that complements the Herschel maps and illustrates the role of ground-based submillimeter telescopes in the identification of candidates for future ALMA studies on the formation of young substellar mass objects. We also highlight the need of multiwavelength studies to interpret properly the results.  

\begin{acknowledgements}
IdG acknowledge support from MICINN (Spain) AYA2011-30228-C03 grant (including FEDER funds). 
HB is funded by the the Ram\'on y Cajal fellowship program number RYC-2009-04497.  
AP acknowledges the financial support from UNAM, and CONACyT (M\'exico).
IdG acknowledges APEX staff for the hard work and support during LABOCA observation.  This publication makes use of data products from the Wide-field Infrared Survey Explorer, Two Micron All Sky Survey, the Spitzer Space Telescope, Herschel and AKARI.
This research has been funded by Spanish grants AYA2012-38897-C02-01 and PRICIT-S2009/ESP-1496.

\end{acknowledgements}

\onecolumn

\begin{appendix}

\section{Counterparts Identification}
\label{method}
For identifying counterparts of the submillimeter cores detected with LABOCA at other wavelengths, we performed an X-matching procedure following the next strategy: We took as base for the X-matching radius the one derived from the LABOCA beam and we proceeded in pairs crossing this LABOCA parental catalog with each catalog corresponding to a different frequency and/or instrument (individual catalogs corresponding to: 2MASS, AKARI, DENIS, Spitzer/IRAC, channels 1-4 independently, Spitzer/MIPS, channels 1 and 2 independently, WFI and VLT/VIMOS; and finally WISE; see Table~\ref{multiwave}). Herschel SPIRE and PACS maps were not used for counterparts identification due to the clumpy nature of its emission. Nevertheless these data were crucial to confirm the detection of the faintest cores observed in our LABOCA map. 

\begin{figure*}[ht!]
\centering
     \includegraphics[width=18.5cm, angle=0]{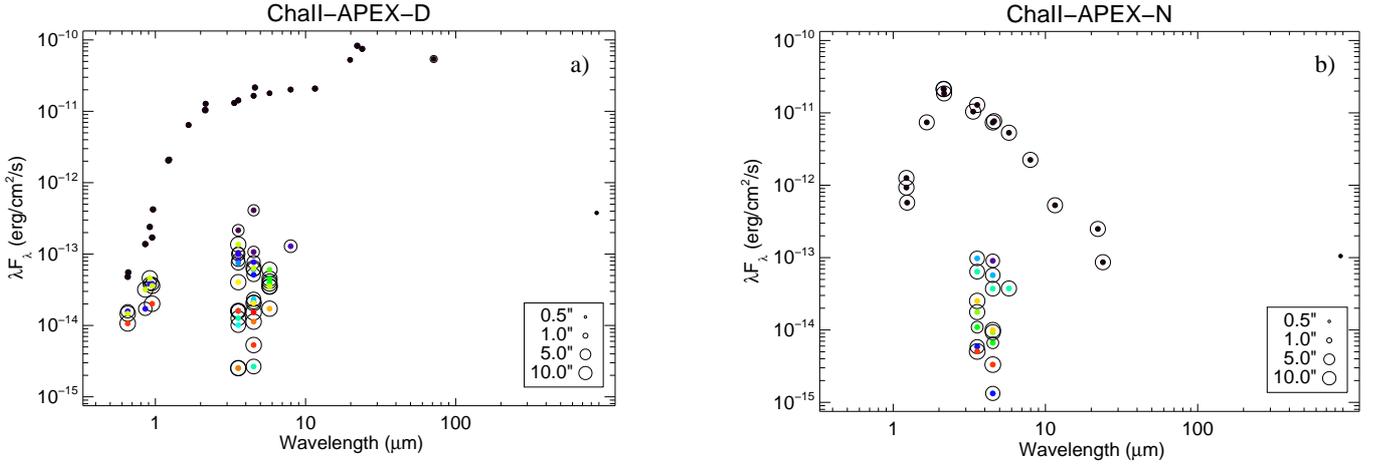}
     \caption{Flux versus wavelength of all the counterparts candidates found for core D (a)) and core N (b)). Dots of different colors represent different counterpart candidates that has the same spatial position (after considering pointing errors). The distance of the counterpart candidates to the nominal position of the submillimeter cores is given by the size of the black circle that surrounds each colored dot.  At the right bottom of each figure (a) and b)) a diagram with the equivalence between black circle radius and the distance to the cores is indicated. For core D, the closest counterpart candidate has a SED typical from a YSO, while the rest of the sources located at other more distant positions show no clear nature. In core N, there is not any YSO associated and only a source with a SED typical from a extincted background star (black dots) and hence not associated with the submillimeter core is found.}
     \label{counterparts}
\end{figure*}

Once we had the individual possible counterparts for each match, we visually inspected those positions in the IRAC, WISE and 2MASS images and discarded all possible counterparts that corresponded to artifacts in the images (areas compromised by saturation, bright star spikes, ghosts, etc.).
For the counterparts that passed the first visual inspection of the images, we performed a combined analysis of the counterparts' astrometry, grouping those objects whose positions at all different
wavelengths are compatible amongst each other given the typical uncertainty in position of those catalogs (much more precise than the LABOCA beam), see Fig~\ref{counterparts}.
%($R = 27.6\sqrt2/2 = 19.52''$) 
%In the final step of the strategy, we discarded counterparts that
%having MIR detections, their MIR slopes were consistent with
%Rayleighâ€“Jeans, and so having no MID excess are not likely to be the
%source of the LABOCA detection.
For revealing the nature of the counterparts found, we studied the SEDs properties using Robitaille's SED tool \citep{Rob06,Rob07}, which allowed us to disentangle between background stars and young stellar objects candidates.

\begin{table*}[ht!]
\small
\caption{\label{multiwave-example} Photometry for the YSO counterpart found for clump D and K, and for the background star spatially coincident with clump N.}
\centering
\begin{tabular}{lcccccccc}
\hline\hline
Telescope/& Filter &  $\lambda$ & S$_{\nu}$ (D)& $\sigma$ (D) & S$_{\nu}$  (N)& $\sigma$ (N) & S$_{\nu}$  (DK Cha) & $\sigma$ (DK Cha) \\
Instrument or survey&  & ($\mu$m) &(Jy) &(Jy) &(Jy) &(Jy) &(Jy) &(Jy)  \\
\hline

ESO 2.2m/WFI                         &R$_{c}$           &0.65 	 &1.05e-05    &5.61e-07    &...           &...	    &7.37e-04   &1.18e-06     \\
ESO 2.2m/WFI                         &H$_{\alpha}$       &0.66 	 &1.22e-05    &1.39e-06    &...           &...	    &1.05e-03   &3.89e-06     \\
ESO 2.2m/WFI                         &NB665		&0.67    &...         &...         &...		  &...	    &6.72e-04   &2.97e-06     \\
ESO 2.2m/WFI                         &MB856             &0.86 	 &3.94e-05    &2.01e-06    &...           &...	    &9.63e-03   &1.21e-05     \\
ESO 2.2m/WFI                         &MB914             &0.92 	 &7.34e-05    &1.71e-06    &...           &...	    &1.03e-02   &1.35e-05     \\
ESO 2.2m/WFI                         &Z                 &0.96 	 &1.36e-04    &6.63e-06    &...           &...	    &1.72e-02   &2.54e-05     \\
ESO 2.2m/WFI                         &I$_{c}$           &0.96 	 &5.41e-05    &1.03e-06    &...           &...	    &6.78e-03   &7.64e-06     \\
ESO 1m/DENIS                         &J                 &1.2     &8.37e-04    &1.47e-04    &5.13e-04    &1.18e-04   &1.27e-01   &5.85e-03     \\
ESO 1m/DENIS                         &K$_{s}$           &2.1 	 &7.45e-03    &8.23e-04    &1.52e-02    &1.26e-03   &5.84       &4.83e-01     \\
ESO 1m/DENIS			     &I			&0.8     &...  	      &...	   &...		&...	    &3.15e-03   &1.16e-04     \\	    
Mt. Hopkins-CTIO/2MASS               &J                 &1.2 	 & 8.69e-04   &6.32e-05    &2.37e-04    &4.58e-05   &2.98e-01   &6.04e-03     \\
Mt. Hopkins-CTIO/2MASS               &H                 &1.7 	 & 3.56e-03   &1.38e-04    &4.10e-03    &1.13e-04   &1.62       &7.02e-02     \\
Mt. Hopkins-CTIO/2MASS               &K$_{s}$           &2.2 	 &9.19e-03    &3.55e-04    &1.34e-02    &3.69e-04   &5.62       &2.43e-01     \\
WISE                                 &W1                &3.4     &1.46e-02    &3.37e-04    &1.16e-02    &2.57e-04   &9.80       &2.80e-01     \\
WISE                                 & W2               &4.6 	 &3.31e-02    &6.40e-04    &1.18e-02    &2.17e-04   &33.73      &6.21e-02     \\
WISE                                 & W3               &12 	 &8.02e-02    &1.26e-03    &2.04e-03    &1.48e-04   &42.74      &7.87e-02     \\
WISE                                 & W4               &22      &6.10e-01    &1.07e-02    &1.83e-03    &1.67e-01   &55.50      &6.13e-01     \\
Spitzer/IRAC                         &  I1              &3.6 	 &1.69e-02    &5.80e-05    &1.53e-02    &5.10e-05   &...	        &...   	     \\
Spitzer/IRAC                         &  I2              &4.5 	 &2.48e-02    &4.40e-05    &1.11e-02    &3.40e-05   &...          &...            \\
Spitzer/IRAC                         &  I3              &5.8     &3.44e-02    &1.17e-04    &1.02e-02    &5.10e-05   &...          &...            \\
Spitzer/IRAC                         &  I4              &8.0     &5.35e-02    &6.80e-05    &5.95e-03    &2.00e-05   &...          &...            \\
Spitzer/MIPS                         &  M1              &24 	 &5.96e-01    &2.95e-04    &6.87e-04    &8.90e-05   &...          &...   	     \\
Spitzer/MIPS                         &  M2              &70 	 &1.29        &9.52e-03    &...	        &...          &...   	&...   	     \\
Akari/IRC			     &S9W		&9.2     &...	      &...	   &...		&...	    &30.07	&1.90        \\
Akari/IRC                            &L18W              &20 	 &3.47e-01    &3.15e-02    &... 	        &...	    &62.52  	&6.27  	     \\
APEX/LABOCA                          &...               &870   	 &1.09e-01    &5.00e-03    &... 	        &... 	    &1.75   	&4.00e-03    \\

\hline
\hline                                                        
\end{tabular}                                                 
\end{table*}

\end{appendix} 


\begin{thebibliography}{}

\bibitem[Alcal{\'a} et al.(2000)]{Alc00} Alcal{\'a}, J.~M., Covino, E., Sterzik, M.~F., et al.\ 2000, \aap, 355, 629 

\bibitem[Alcal{\'a} et al.(2008)]{Alc08} Alcal{\'a}, J.~M., Spezzi, L., Chapman, N., et al.\ 2008, \apj, 676, 427

\bibitem[Andr{\'e} et al.(2010)]{And10} Andr{\'e}, P., Men'shchikov, A., Bontemps, S., et al.\ 2010, \aap, 518, L102 

\bibitem[Andr{\'e} et al.(1993)]{And93} Andre, P., Ward-Thompson, D., \& Barsony, M.\ 1993, \apj, 406, 122 

\bibitem[Andr{\'e} et al.(2000)]{And00} Andr{\'e}, P., Ward-Thompson, D., \& Barsony, M.\ 2000, Protostars and Planets IV, 59 

\bibitem[Andr{\'e} et al.(2012)]{And12} Andr{\'e}, P., Ward-Thompson, D., \& Greaves, J.\ 2012, Science, 337, 69 

\bibitem[Allen et al.(2004)]{All04} Allen, L.~E., Calvet, N., D'Alessio, P., et al.\ 2004, \apjs, 154, 363 

\bibitem[Barrado y Navascu{\'e}s \& Jayawardhana(2004)]{Bar04} Barrado y Navascu{\'e}s, D., \& Jayawardhana, R.\ 2004, \apj, 615, 840 

\bibitem[Barrado et al.(2009)]{Bar09} Barrado, D., Morales-Calder{\'o}n, M., Palau, A., et al.\ 2009, \aap, 508, 859 

\bibitem[Bate et al.(2002)]{Bat02} Bate, M.~R., Bonnell, I.~A., \& Bromm, V.\ 2002, \mnras, 332, L65 

\bibitem[Bate(2012)]{Bat12} Bate, M.~R.\ 2012, \mnras, 419, 3115 

\bibitem[Bonnor(1956)]{Bon56} Bonnor, W.~B.\ 1956, \mnras, 116, 351 

\bibitem[Epchtein et al.(1997)]{Epc97} Epchtein, N., de Batz, B., Capoani, L., et al.\ 1997, The Messenger, 87, 27 

\bibitem[Evans et al.(2003)]{Eva03} Evans, N.~J., II, Allen, L.~E., Blake, G.~A., et al.\ 2003, \pasp, 115, 965 

\bibitem[Evans et al.(2001)]{Eva01} Evans, N.~J., II, Rawlings, J.~M.~C., Shirley, Y.~L., \& Mundy, L.~G.\ 2001, \apj, 557, 193 

\bibitem[Fuente et al.(2007)]{Fue07} Fuente, A., Ceccarelli, C., Neri, R., et al.\ 2007, \aap, 468, L37 

\bibitem[Greaves et al.(2003)]{Gre03} Greaves, J.~S., Holland, W.~S., \& Pound, M.~W.\ 2003, \mnras, 346, 441 

\bibitem[Gueth et al.(2003)]{Gue03} Gueth, F., Bachiller, R., \& Tafalla, M.\ 2003, \aap, 401, L5 

\bibitem[Hacar et al.(2013)]{Hac13} Hacar, A., Tafalla, M., Kauffmann, J., \& Kov{\'a}cs, A.\ 2013, \aap, 554, A55 

\bibitem[Hartigan(1993)]{Har93} Hartigan, P.\ 1993, \aj, 105, 1511 

\bibitem[K{\"o}nyves et al.(2010)]{Kon10} K{\"o}nyves, V., Andr{\'e}, P., Men'shchikov, A., et al.\ 2010, \aap, 518, L106 

\bibitem[Kov{\'a}cs(2008)]{Kovacs08} Kov{\'a}cs, A.\ 2008, \procspie, 7020,  

\bibitem[Lada(1987)]{Lad87} Lada, C.~J.\ 1987, Star Forming Regions, 115, 1 

\bibitem[Larson et al.(1998)]{Lar98} Larson, K.~A., Whittet, D.~C.~B., Prusti, T., \& Chiar, J.~E.\ 1998, \aap, 337, 465 

\bibitem[Lee et al.(2013)]{Lee13} Lee, C.~W., Kim, M.-R., Kim, G., et al.\ 2013, \apj, 777, 50 

\bibitem[Lim \& Tan(2014)]{Lim14} Lim, W., \& Tan, J.~C.\ 2014, \apjl, 780, L29 

  
\bibitem[Maury et al.(2010)]{Mau10} Maury, A.~J., Andr{\'e}, P., Hennebelle, P., et al.\ 2010, \aap, 512, A40 

\bibitem[Ossenkopf \& Henning(1994)]{Ose94} Ossenkopf, V., \& Henning, T.\ 1994, A\&A, 291, 943 

\bibitem[Padoan \& Nordlund(2004)]{Pad04} Padoan, P., \& Nordlund, {\AA}.\ 2004, \apj, 617, 559 

\bibitem[Palau et al.(2012)]{Pal12} Palau A., de Gregorio-Monsalvo, I., Morata, O., Stamatellos, D., Hu\'elamo, N., Eiroa, C., Bayo, A., Morales-Calderon, M., Bouy, H., Ribas, A., Asmus, D., Barrado, D. \ 2012, \mnras, 424, 2778.

\bibitem[Palau et al.(2014)]{Pal14} Palau, A., Zapata, L. A., Rodr\'iguez, L. F. et al. 2014, MNRAS, submitted 

\bibitem[Persi et al.(2003)]{Per03} Persi, P., Marenzi, A.~R., G{\'o}mez, M., \& Olofsson, G.\ 2003, \aap, 399, 995 

\bibitem[Rathborne et al.(2005)]{Rat05} Rathborne, J.~M., Jackson, J.~M., Chambers, E.~T., et al.\ 2005, \apjl, 630, L181 

\bibitem[Reipurth \& Clarke(2001)]{Rei01} Reipurth, B., \& Clarke, C.\ 2001, \aj, 122, 432 

\bibitem[Robitaille et al.(2006)]{Rob06} Robitaille, T.~P., Whitney, B.~A., Indebetouw, R., Wood, K., \& Denzmore, P.\ 2006, \apjs, 167, 256 

\bibitem[Robitaille et al.(2007)]{Rob07} Robitaille, T.~P., Whitney, B.~A., Indebetouw, R., \& Wood, K.\ 2007, \apjs, 169, 328

\bibitem[Roy et al.(2013)]{Roy13} Roy, A., Martin, P.~G., Polychroni, D., et al.\ 2013, \apj, 763, 55 

\bibitem[Schwartz(1977)]{Sch77} Schwartz, R.~D.\ 1977, \apjs, 35, 161 

\bibitem[Spezzi et al.(2008)]{Spe08} Spezzi, L., Alcal{\'a}, J.~M., Covino, E., et al.\ 2008, \apj, 680, 1295

\bibitem[Spezzi et al.(2013)]{Spe13} Spezzi, L., Cox, N.~L.~J., Prusti, T., et al.\ 2013, \aap, 555, A71 

\bibitem[Stamatellos \& Whitworth(2009)]{Sta09} Stamatellos, D., \& Whitworth, A.~P.\ 2009, \mnras, 392, 413 

\bibitem[van Kempen et al.(2010)]{Van10} van Kempen, T.~A., Green, J.~D., Evans, N.~J., et al.\ 2010, \aap, 518, L128 

\bibitem[Vuong et al.(2001)]{Vuo01} Vuong, M.~H., Cambr{\'e}sy, L., \& Epchtein, N.\ 2001, \aap, 379, 208 

\bibitem[Wagle et al.(2015)]{Wag15} Wagle, G.~A., Troland, T.~H., Ferland, G.~J., \& Abel, N.~P.\ 2015, \apj, 809, 17 

  
\bibitem[Whittet et al.(1997)]{Whi97} Whittet, D.~C.~B., Prusti, T., Franco, G.~A.~P., et al.\ 1997, \aap, 327, 1194 

\bibitem[Whitworth \& Zinnecker(2004)]{Whi04} Whitworth, A.~P., \& Zinnecker, H.\ 2004, \aap, 427, 299 

\end{thebibliography}
\end{document}